\documentclass[pre,twocolumn,amsmath,amssymb,superscriptaddress,showpacs]{revtex4-2}
\usepackage{graphicx}
\usepackage{latexsym}
\usepackage{dcolumn}
\usepackage{bm}
\usepackage{color}
\usepackage{amssymb}
\usepackage[normalem]{ulem}
\usepackage{mathtools}

\newcommand{\be}{\begin{equation}}
\newcommand{\ee}{\end{equation}}
\newcommand{\bea}{\begin{eqnarray}}
\newcommand{\eea}{\end{eqnarray}}
\newcommand{\bse}{\begin{subequations}}
\newcommand{\ese}{\end{subequations}}

\newcommand{\erf}{\mbox{erf}}

\newcommand{\red}[1]{\textcolor{black}{#1}}
\newcommand{\e}{\epsilon}

\newcommand{\comment}[1]{}

\begin{document}

\title{Condensation induced by coupled transport processes}
\date{\today}

\author{Gabriele Gotti}
\affiliation{Dipartimento di Fisica e Astronomia, Universit\`a di Firenze,
via G. Sansone 1 I-50019, Sesto Fiorentino, Italy}
\affiliation{Istituto dei Sistemi Complessi, Consiglio Nazionale
delle Ricerche, via Madonna del Piano 10, I-50019 Sesto Fiorentino, Italy}

\author{Stefano Iubini}
\email{stefano.iubini@cnr.it}
\affiliation{Istituto dei Sistemi Complessi, Consiglio Nazionale
delle Ricerche, via Madonna del Piano 10, I-50019 Sesto Fiorentino, Italy}
\affiliation{Istituto Nazionale di Fisica Nucleare, Sezione di Firenze, 
via G. Sansone 1 I-50019, Sesto Fiorentino, Italy}

\author{Paolo Politi}
\email{paolo.politi@cnr.it}
\affiliation{Istituto dei Sistemi Complessi, Consiglio Nazionale
delle Ricerche, via Madonna del Piano 10, I-50019 Sesto Fiorentino, Italy}
\affiliation{Istituto Nazionale di Fisica Nucleare, Sezione di Firenze, 
via G. Sansone 1 I-50019, Sesto Fiorentino, Italy}

\begin{abstract}
Several lattice models display a condensation transition in real space
when the density of a suitable order parameter exceeds a critical value.
We consider one of such models with two conservation laws,
in a one-dimensional open setup where the system is attached to two external
reservoirs.
\red{Both reservoirs impose subcritical boundary conditions at the chain ends.
When such boundary conditions are equal,
the system is in
equilibrium below the condensation threshold and no condensate can appear.
Instead, when the system is kept out of equilibrium,
localization may arise in an internal portion of the lattice.}
We discuss the origin of this phenomenon, the relevance of the number 
of conservation laws, and the effect of the pinning of the condensate 
on the dynamics of the out-of-equilibrium state.
\end{abstract}
\maketitle

\section{Introduction}
Real-space condensation phenomena represent a vast class of processes that appear in many domains of
physics.  Typical examples from the classical realm 
are the behavior of traffic flows, the dynamics of shaken granular media, the aggregation and fragmentation
processes, the localization in
real-world networks, phase transitions in stochastic mass transport models, nonlinear propagation in
weakly dissipative systems~%
\cite{Majumdar1998_PRL,eggers99,Schutz2003_JPA,Kafri2002_PRL,Godreche2003_JPA,Evans2005_JPA,%
Majumdar2010_LesHouches,torok05,Zannetti2014_PRE,Pastor2016_SR,Eisenberg1998,Lederer2008}.
All these systems may display a phase transition from a homogeneous phase to a localized one 
when varying a suitable control parameter, which is often the density of a conserved quantity of the system: 
above a critical density
a finite fraction of the conserved quantity is localized on a single lattice site, while
 the remaining amount is delocalized over the rest of the system~%
\cite{Drouffe1998_JPA,Majumdar2005_PRL,Godreche2007_LNP,Godreche2021,GIP21}.
In the absence of intrinsic dishomogeneities such as spatial disorder~\cite{Godreche2012_JSTAT,Barre2018_JSTAT}, 
each lattice site is equally eligible to host the condensate.

\red{So far, the condensation transition has been studied in details in a large variety of equilibrium
models and in translationally invariant, bulk-driven diffusive models~\cite{KK_JPA_pedestrian}.
In the latter case, the nonequilibrium condition is ascribable to the presence of a bias
in the microscopic dynamics which breaks detailed balance in a system
with periodic boundary conditions. Hence macroscopic circulating currents are steadily induced.
In some cases, e.g. in the Zero-Range Processes (ZRP)~\cite{Evans2005_JPA}, the bias breaks 
detailed balance but the nonequilibrium steady state equals the equilibrium state of the same
model when hopping is made symmetric. 
}

\red{In this paper we intend to analyze a boundary-driven model where
translational invariance is broken and which is genuinely off-equilibrium
because the system is in contact with two different thermal reservoirs at its ends.
In this setup bulk dynamics is assumed to be reversible, local equilibrium
may appear, and conservation laws are relevant for the dynamics.
Condensation in the context of such an open, out-of-equilibrium  system 
has been much less explored. 
Asymmetric boundary conditions have been studied for ZRP,
i.e. in models with a single conservation law, but in this case condensation appears
only if a boundary condition is overcritical~\cite{Levine2005}, i.e. if a reservoir imposes a localized equilibrium state.
}

\begin{figure}[t]
\begin{center}
\includegraphics[width=0.46\textwidth,clip]{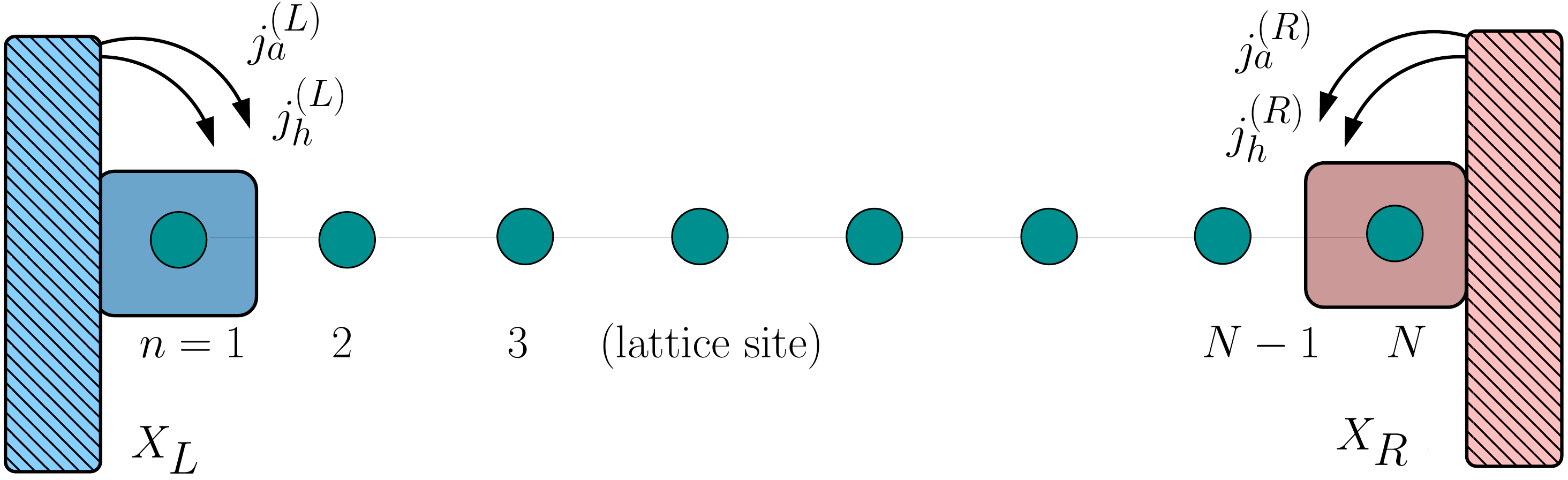}
\end{center}
\caption{\red{The setup: a one-dimensional lattice with $N$ sites interacts  with two external reservoirs at its boundaries.
$X_L$ and $X_R$ are the thermodynamic parameters imposed at the left and right side, respectively. Each reservoir exchanges 
with the system two distinct quantities (mass $a$ and energy $h$, see text), which are transported through the chain.  
The corresponding fluxes from the reservoirs to the system are denoted by $j_{a,h}^{(L,R)}$, see Sec.~\ref{sec:model}.
In our case $X_L$ and $X_R$ are such that left currents are negative and right currents are positive.
In a steady state there is a perfect balance between them, $j^{(L)} + j^{(R)} =0$.
}}
\label{fig:setup}
\end{figure}

\red{
In this paper 
we provide evidence that a new condensation mechanism may arise in out-of-equilibrium conditions but this
occurs only if the system has two conservation laws.
More precisely, we focus on a setup in which a one-dimensional 
lattice chain with local interactions and two conserved quantities is in contact with external reservoirs
imposing thermodynamic parameters $X_L$ and $X_R$ 
at its left and right boundaries, respectively, see Fig.~\ref{fig:setup}. 
Importantly, we consider the case in which both $X_L$ and $X_R$  are in the homogeneous, delocalized phase.
With this constraint, in equilibrium conditions ($X_L=X_R$) condensation is always forbidden.
On the other hand, we find that when $X_L \neq X_R$ localization may arise in the bulk of the system.
}

As we will see, the existence of two conservation laws implies 
a coupled transport process which is essential to induce condensation.
We will also show that the properties of the asymptotic, out-of-equilibrium state depend
on the possibility that the condensate may diffuse or not in space.
If diffusion is allowed, the system attains a nonequilibrium steady state;
if the condensate is pinned, it does not stop growing  and stationarity is lost.

\red{
The paper is organized as follows. In Sec.~\ref{sec:model} we introduce the model and discuss its equilibrium properties.
In Sec.~\ref{sec:noneq} we present the phenomenon of condensation occurring in a coupled-transport setup and we separately analyze
the case of pinned and unpinned dynamics. Finally, in Sec.~\ref{sec:disc} we provide a discussion of the main results and 
in Sec.~\ref{sec.perspectives} we outline the major open problems and perspectives.
} 

\section{The Model and its equilibrium properties}\label{sec:model}
\begin{figure}
\begin{center}
\includegraphics[width=0.47\textwidth,clip]{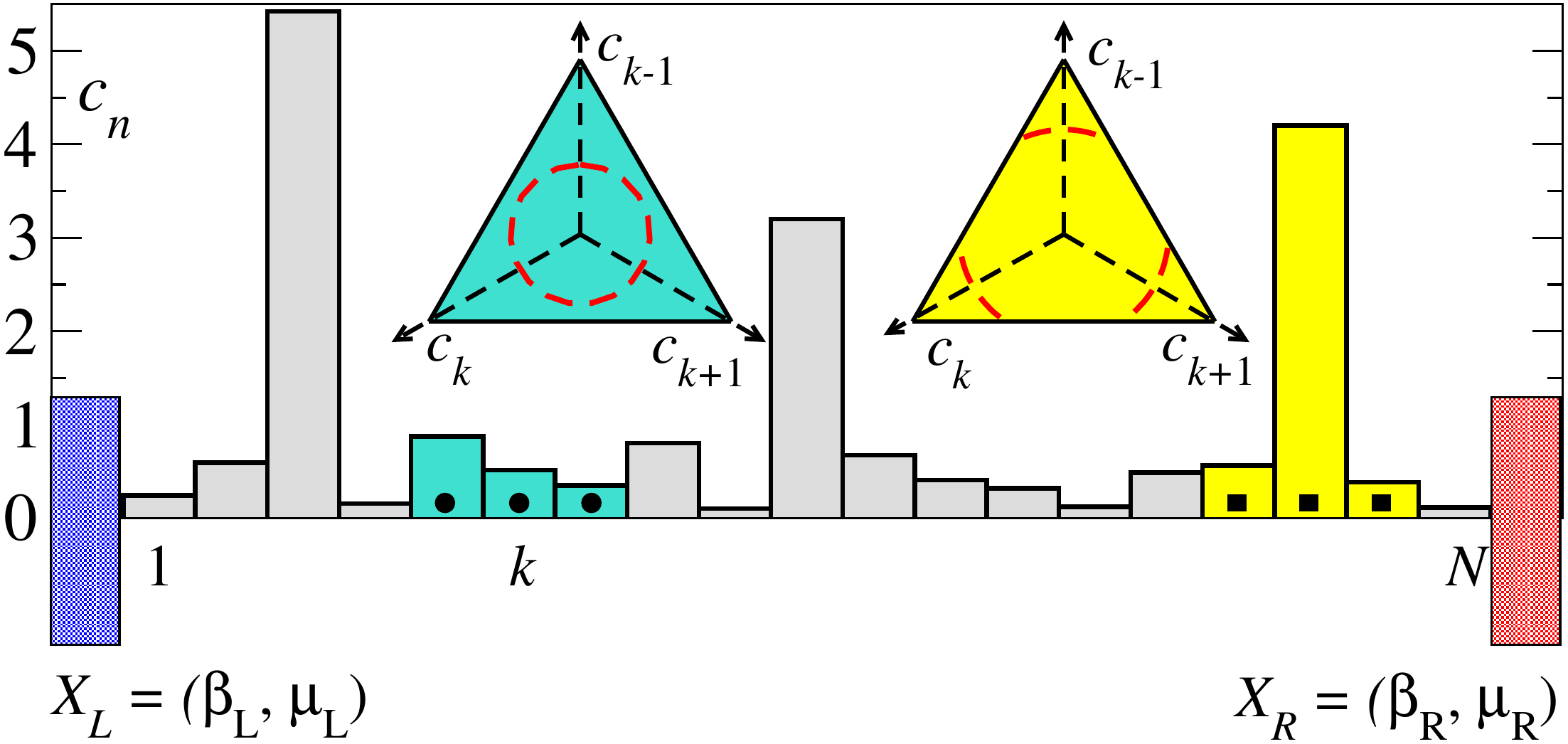}
\end{center}
\caption{
Sketch of the C2C model with open boundary conditions: the two reservoirs 
interact at the boundaries 
of the chain and \red{impose thermal parameters $X_L=(\beta_L,\mu_L)$ and $X_R=(\beta_R,\mu_R)$}. The bulk dynamics is realized by implementing the MMC algorithm (see text) on random triplets of consecutive
sites. Conservation of energy and mass  restricts the available states in the space $(c_{k-1},c_k,c_{k+1})$ either on a circle or on the union of three distinct
arcs (red dashed lines).
}
\label{fig.c2c}
\end{figure}

The model we are going to describe can be derived from
the Discrete Nonlinear Schr\"odinger (DNLS) equation~\cite{kevrekidis09}, 
which is ubiquitous in nonlinear physics 
with  applications in optics, cold-atoms physics and micro-magnetic systems. 
\red{In its standard form, the DNLS equation reads
\be
i\dot{z}_n = -2|z_n|^2 z_n - z_{n+1} -z_{n-1}
\ee
where $z_n$ are complex-valued amplitudes defined on a one-dimensional lattice with $1\leq n \leq N$.
This equation displays two exactly conserved quantities, namely the total energy
\be\label{eq:hdnls}
H=\sum_n |z_n|^4 + z^*_n z_{n+1} + z_n z^*_{n+1}
\ee
 and the total norm of the wavefunction 
\be
A=\sum_n |z_n|^2\,,
\ee 
also called total ``mass"~\cite{Rasmussen2000_PRL}.} For this reason, the model displays
coupled transport when it is steadily maintained out of equilibrium~\cite{iubini12,Iubini2016_NJP}. 
Above a certain critical energy density which depends quadratically on the mass density, the DNLS model also exhibits
a condensation transition with a very rich phenomenology that 
involves the excitation of {\it discrete breather} states~\cite{Rasmussen2000_PRL,Iubini2013_NJP}
 and
includes the creation of 
negative-temperature states~\cite{Iubini2013_NJP},
extremely slow thermalization~\cite{PPI2022_JSTAT}, 
inequivalence of statistical ensembles~\cite{Gradenigo2021_JSTAT}, 
and ergodicity breaking~\cite{Flach2018_PRL}.
Its equilibrium picture has recently reached a systematic and accurate description within the 
so-called large deviation theory~\cite{Gradenigo2021_JSTAT,Mori2021}. 
This has been possible because near the transition the DNLS model simplifies, 
\red{
as the hopping energy $(z_n^*z_{n+1} +c.c.)$ in Eq.~(\ref{eq:hdnls})
becomes negligible with respect to the local energy $|z_n|^4$~\cite{Gradenigo2021_JSTAT}.
The resulting simplified model, which is the object of our study, can be therefore expressed in terms
of the local masses $c_n = |z_n|^2 \ge 0$ only. 
We refer to this model as C2C~\cite{GIP21}, see Fig.~\ref{fig.c2c}, because it displays
condensation in the presence of two conserved quantities, the mass $A$ and the energy $H$,}
\bea\label{eq:AH}
A &=& \sum_{n=1}^N c_n\,,\\ 
H &=& \sum_{n=1}^N c_n^2\,.
\eea

\begin{figure}
\begin{center}
\includegraphics[width=0.49\textwidth,clip]{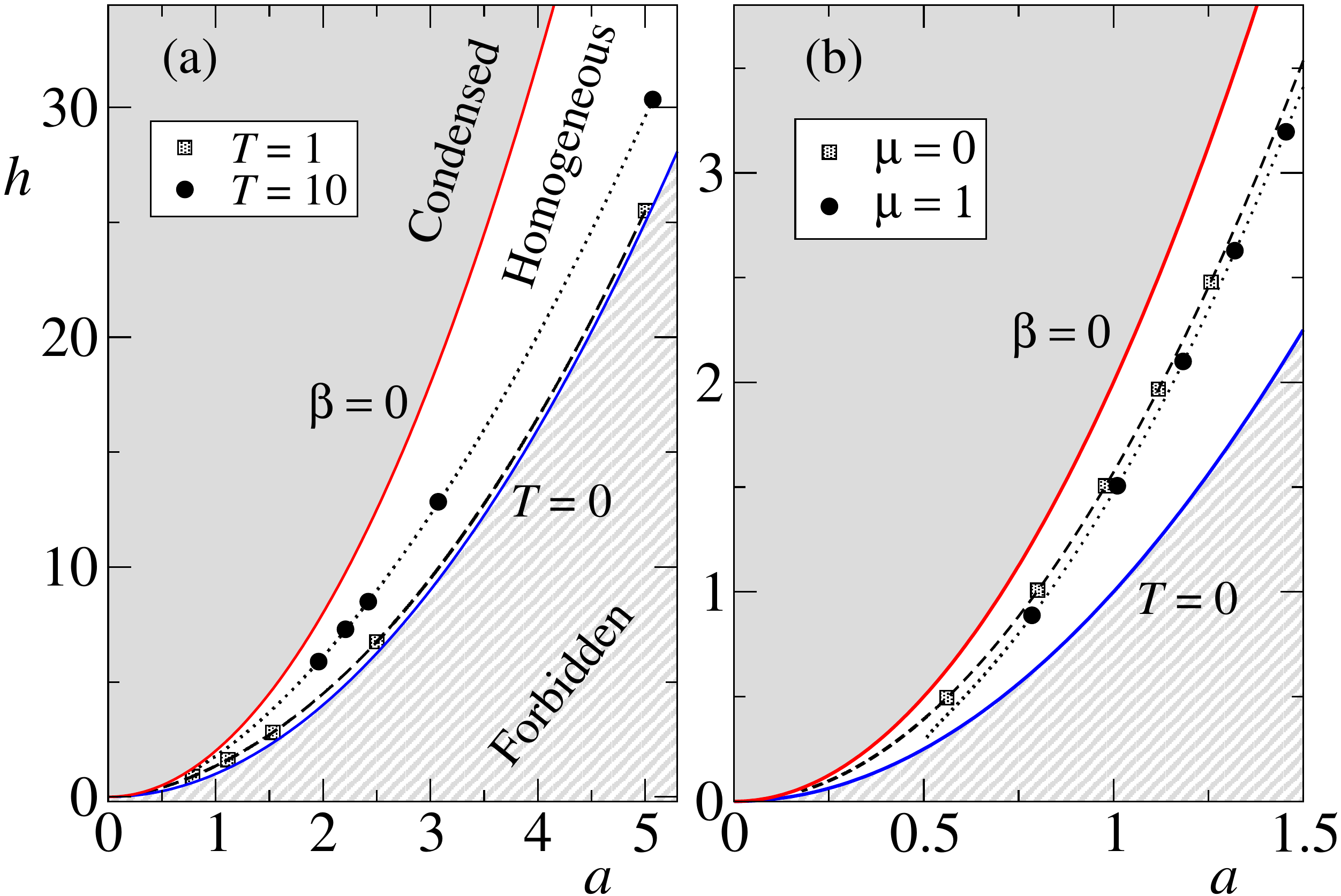}
\end{center}
\caption{
Equilibrium phase diagram $(a,h)$: red and blue solid lines identify the critical line $\beta=0$ and the ground state
$T=0$, respectively. Heat baths are defined in the homogeneous region and are consistent with the grand canonical 
description of Eqs.~(\ref{eq.a_gc}-\ref{eq.h_gc}). Here we compare the values of $a$ and $h$ obtained from 
grand canonical equilibrium
simulations (full symbols) with the analytical curves relating $(a,h)$ to $(\beta,\mu)$ (continuous lines).
Panels (a) and (b) show isothermals ($T=$const) and isochemical ($\mu=$const) lines, respectively. 
}
\label{fig.ahTmu}
\end{figure}

\red{
The thermodynamic behavior of the C2C model in equilibrium is specified by the values of the mass density $a=H/N$ and
of the energy density $h=H/N$. Therefore one can profitably represent all equilibrium states in the plane $(a,h)$, see Fig~\ref{fig.ahTmu}.}
\red{In order to identify the main features of the equilibrium diagram,} 
it is useful to evaluate the variance of the mass distribution,
\be
\sigma^2 = \overline{c^2} - {\overline{c}}^2 = h - a^2, 
\label{eq.sigma}
\ee
which shows that it must be $h\ge a^2$. For $h=a^2$, $c_n \equiv a$ and $\sigma=0$.
With increasing $h$ (at fixed $a$) the variance of the mass distribution increases
until the standard variation of the (positive) mass equals its average value.
This occurs at the critical line $h=2a^2$, where the mass is distributed exponentially~\cite{Rasmussen2000_PRL}.
Above this line a condensation process occurs and in the thermodynamic limit 
a single site hosts the extra-energy $(h-2a^2)N$.

\red{A more detailed description of equilibrium states is obtained from the evaluation of 
the grand canonical partition function~\cite{Szavits2014_PRL,Szavits2014_JPA} derived in App.~\ref{app:gcan},
\be
Z(\beta,\mu) = \left[\int_0^{+\infty}dc\, e^{-\beta( c^2 -\mu c)}\right]^N\,,
\ee
where $\beta$ and $\mu$ are the inverse temperature and the chemical potential, respectively.  
  }
Using the standard formulas $a=\langle A\rangle/N$ and $h=\langle H\rangle/N$, \red{where the symbol $\langle\cdot\rangle$ denotes the average
over the grand canonical distribution}, we find the equations
\bea
\label{eq.a_gc}
a &=& \frac{\mu}{2} + \frac{1}{\sqrt{\pi\beta}} 
\frac{e^{-\beta\mu^2/4}}{1 +\erf\left(\frac{\sqrt{\beta}\mu}{2}\right)} \\
h &=& \frac{1}{2\beta} +\frac{1}{2}a\mu ,
\label{eq.h_gc}
\eea
which allow to connect $(a,h)$ to $(\beta,\mu)$ and vice versa. \red{The details of the
calculation are reported in  Appendix~\ref{app:gcan}}.
In particular it is found that the critical parabola $h=2a^2$ corresponds to
$T\equiv1/\beta=+\infty$ and $\beta\mu=-1/a$, while the parabola $h=a^2$ corresponds to 
$T=0$ and $\mu=2a$. States below the $T=0$ line are forbidden and Eqs.~(\ref{eq.a_gc}-\ref{eq.h_gc})
can not account for the states above the $T=+\infty$ line. 
In fact, as proven in Refs.~\cite{Gradenigo2021_JSTAT,Gradenigo2021_EPJE}, 
the grand canonical ensemble is not able to describe the
localized phase $h>2a^2$.

The microscopic evolution of the C2C model is implemented numerically with a  
 microcanonical Monte Carlo (MMC) algorithm~\cite{JSP_DNLS,JSTAT_mmc}.
According to this algorithm, random updates of the $c_n$ are identified by
the simplest moves that conserve mass and
energy, simultaneously. Since two conservation laws imply two constraints
we use a one-parameter evolution rule involving three sites, which we assume to be contiguous: $(n-1,n,n+1)$. 
In practice we pick up randomly a site $n$ 
and we make a uniform choice of the new masses $(c'_{n-1},c'_n,c'_{n+1})$ such that:
\red{{\it i)} the total mass and total energy of the triplet are conserved and {\it ii)} the detailed balance is 
satisfied~\cite{JSP_DNLS}.
As explained in Appendix~\ref{app:mmc} }
this amounts
to choose a random point in the intersection between a plane and a sphere,
see the red dashed lines in Fig.~\ref{fig.c2c}.
Because of the constraint $c_n \ge 0$ such curve is either a circle or
the union of three disjoint arcs, the latter case occurring when
the mass of one of these sites is significantly larger than the other masses.
\red{Different arcs correspond to the same triplets where there has been a
permutation of the sites such that the position of the peak has moved.}

\red{
When the intersection of plane and sphere is given by three disjoint arcs we have two possible choices
for the dynamics:
either the new triplet of masses is forced to belong to the same arc 
containing the initial state or 
it is allowed to stay in any of the three arcs.
In the former ``pinned" case, high energy peaks 
cannot move to the neighboring sites, while in the latter case breathers
can freely diffuse  in the lattice (``unpinned" case). 
These two different update rules are equivalent in equilibrium conditions~\cite{JSP_DNLS} but they become
inequivalent out of equilibrium.}

When the C2C chain is attached to external reservoirs  as in Fig.~\ref{fig.c2c},
the evolution proceeds by choosing a random site $n=1,\dots,N$ and distinguishing between
inner $(n=2,\dots,N-1)$ and outer $(n=1,N)$ sites.
For inner sites the triplet $(n-1,n,+1)$ belongs to the chain and evolution proceeds as for the MMC
algorithm.
For outer sites we perform a standard grand canonical Monte Carlo  move according to given thermodynamic parameters
\red{$X_L\equiv(\beta_L,\mu_L)$ (left boundary) and $X_R\equiv (\beta_R,\mu_R)$ (right boundary). 
The details of reservoir dynamics are discussed  in Appendix~\ref{app:mmc}.
The rate of exchange of mass and energy from  the reservoirs to the system can be measured by means
of suitable definitions of mass- and energy fluxes. For the left boundary, we define
\begin{eqnarray}\label{eq:flux}
j_a^{(L)}&=& \frac{1}{\tau} \sum_{t_k=1}^{\tau} \delta c_1(t_k)\nonumber\\
j_h^{(L)}&=& \frac{1}{\tau} \sum_{t_k=1}^{\tau} \delta c_1^2(t_k) \quad \tau\gg1\,,
\end{eqnarray}
where $\delta c_1(t_k)$ and $\delta c_1^2(t_k)$ represent respectively the variations  of mass and energy
on the first lattice site produced by a Monte Carlo move of the reservoir occurring at time $t_k$. 
The definitions of $j_a^{(R)}$ and $j_h^{(R)}$ on the right boundary are readily obtained by replacing $L\rightarrow R$ and $c_1 \rightarrow c_N$
in Eq.~(\ref{eq:flux})
}

\red{When the external heat baths impose the same thermodynamic parameters at the chain ends, $X_L=X_R$,
the system reaches thermodynamic equilibrium and $j_a^{(L,R)}=j_h^{(L,R)}=0$. In this condition,}
 mass- and energy densities fluctuate with respect to average values 
$\bar c$ and $\overline{c^2}$,
respectively.  
If heat baths work properly, we expect that
$\bar c=a$ and $\overline{c^2}=h$,
\red{ where $a$ and $h$  are related to the imposed values $(\beta,\mu)$
  through Eqs.~(\ref{eq.a_gc}-\ref{eq.h_gc})}.
In Fig.~\ref{fig.ahTmu} we show that this is indeed the case.
\red{In detail, isothermal ($T=$ const., left panel) and isochemical $(\mu=$ const., right panel) lines
are here reproduced in the phase diagram $(a,h)$. A very good correspondence is found between numerical simulations
and analytic predictions.}

\section{Localization in out-of-equilibrium conditions}\label{sec:noneq}
\red{In this Section we study the C2C model in out-of-equilibrium conditions ($X_L\neq X_R$) and discuss 
the emergence of localization.}
The left heat bath  (attached to the site $n=1$) 
is defined by \red{parameters} $\mu_L = 1$ and $T_L=0.1$, which correspond
to an average mass $a_1 =0.48$ and an average energy $h_1=0.28$.
The right heat bath (attached to the site $n=N$) 
is defined by $\mu_R=5$ and $T_R=5$, which correspond
to an average mass $a_N =2.69$ and an average energy $h_N=9.22$.
\red{As anticipated, both boundary conditions are subcritical, i.e. $h_1 < 2a_1^2$ and $h_N < 2a_N^2$.
In Figs.~\ref{fig.c2c_pinned} and \ref{fig.c2c_unpinned_1} we  highlight their position in the
phase diagram with filled black circles.
Initial conditions of the C2C chain are always chosen to be homogeneous and to
connect the boundary values of $a_1$ and $a_N$.}

After a sufficiently long relaxation time $\tau_0\gtrsim 10^7$, we perform a time average 
of \red{ the instantaneous} mass  and energy  on each lattice site,
\bea
a_n &=& \frac{1}{\tau} \int_{\tau_0}^{\tau_0+\tau} c_n(t) dt \\
h_n &=& \frac{1}{\tau} \int_{\tau_0}^{\tau_0+\tau} c_n^2(t) dt 
\eea
obtaining the \red{corresponding time-average spatial profiles, $a_n$ and $h_n$.} 
If we plot $h_n(a_n)$ we obtain a parametric curve in the space $(a,h)$
which connects the circles corresponding to the \red{thermal boundary conditions}.
In Fig.~\ref{fig.c2c_pinned} we report our results for pinned dynamics
(localized peaks can not diffuse in the lattice) while in Fig.~\ref{fig.c2c_unpinned_1} we show the results
for the unpinned case (peaks freely diffuse). The main result is that in both cases the parametric curve
enters the localization region $h>2a^2$ \red{and a condensate is spontaneously created. Nevertheless, the condensation scenarios arising from these two dynamical rules
are completely different.}

\subsection{Dynamics with pinning}
\label{sec.pinning}

\begin{figure}
\begin{center}		
\includegraphics[width=0.45\textwidth,clip]{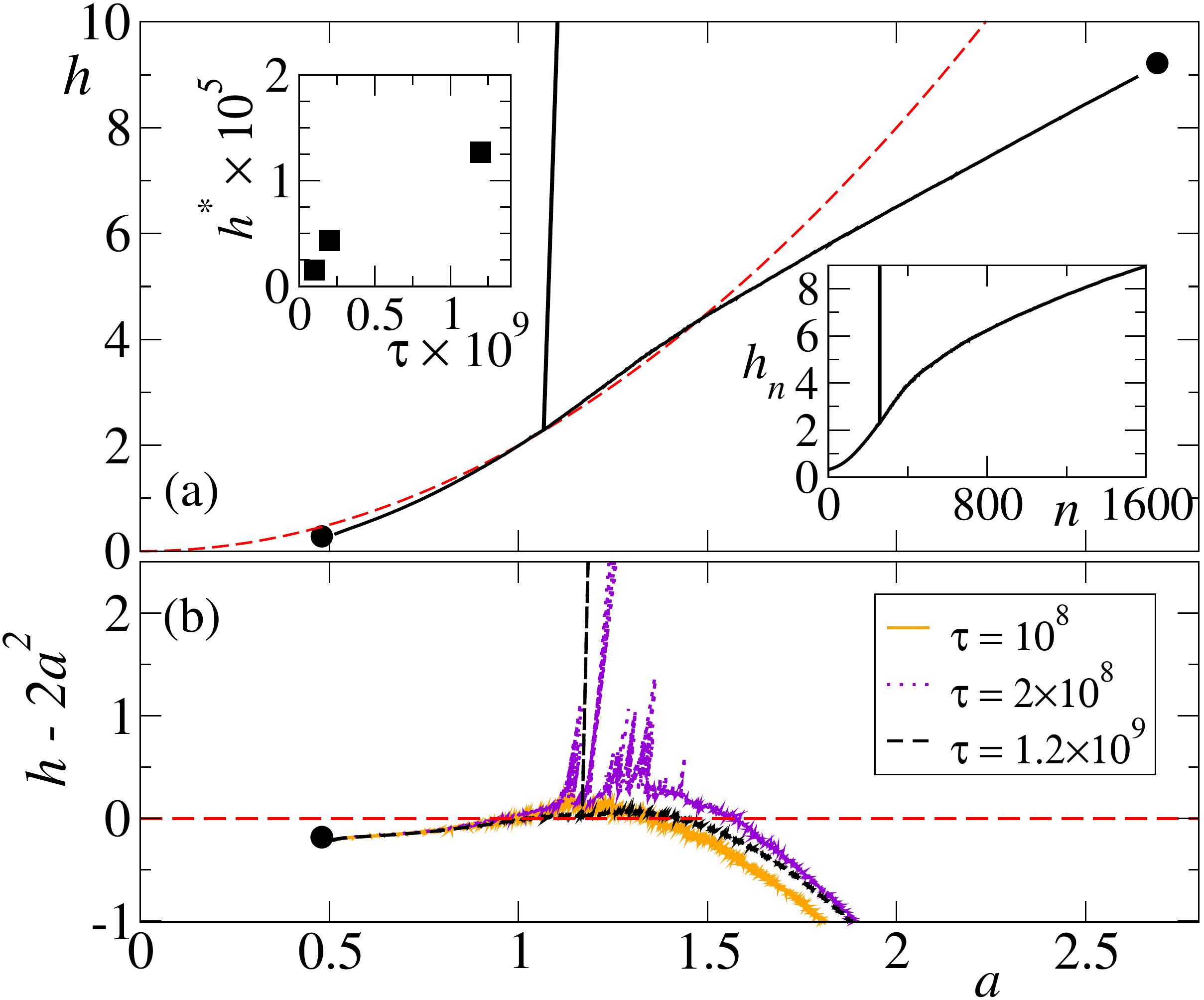}
\end{center}
\caption{
Pinned dynamics. The red dashed lines correspond to the critical line $h=2a^2$.
(a)~Parametric profile in the $(a,h)$ plane for $N=1600$ and simulation time $\tau=1.2\times 10^9$.
In the bottom right inset we plot the corresponding spatial energy profile  $h_n$.
Top left inset reports the peak's energy $h^*$ as a function of time.
(b)~Parametric profiles for $N=12800$ at different times (see legend). \red{The vertical axis reports the quantity $h-2a^2$.}
}
\label{fig.c2c_pinned}
\end{figure}

When pinning is imposed, \red{a typical localized parametric profile is reported in the $(a,h)$ plane in
 the main panel of Fig.~\ref{fig.c2c_pinned}(a) for $N=1600$ and a large simulation time $\tau=1.2\times 10^9$.
  The large peak close to the critical line $h=2a^2$
 corresponds to  a condensate localized on a single lattice site. The related energy density profile $h_n$
 is reported in the lower inset and shows that the  nonlocalized portion of the chain gives rise to a smooth curve.}
\red{In the presence of pinning,} a large condensate localized on a lattice site cannot effectively jump
on neighboring ones~\cite{JSP_DNLS}. This property is found to hinder the relaxation to a  
nonequilibrium  steady state, as the condensate absorbs a fraction of the energy flux transported through the chain
(the effect on the mass flux is negligible because the mass is not condensed, energy is). As a result, the peak's energy $h^*$
increases in time, as shown in the upper inset of Fig.~\ref{fig.c2c_pinned}(a) and boundary fluxes are 
unbalanced, \red{i.e. $j_a^{(L)}\neq -j_a^{(R)}$ and $j_h^{(L)}\neq -j_h^{(R)}$.
The resulting nonequilibrium process is therefore non stationary.} 
\red{The spontaneous localization process}  is found to be stable against size variation: in Fig.~\ref{fig.c2c_pinned}(b)
we show the parametric profiles for a larger chain ($N=12800$) at different simulation times.
\red{For more clarity, we plot on the vertical axis the quantity $h-2a^2$: in this representation the critical line $h=2a^2$ maps to a 
horizontal line passing from the origin.
Condensation starts when a smooth profile approaches and crosses the critical line, see solid orange curve. At intermediate times,
the profile enters the localized region  and a  few localized states arise (dotted purple line). Finally, at larger times,
a single condensed site remains (black short-dashed line), similarly to the case $N=1600$ shown in Fig.~\ref{fig.c2c_pinned}(a).}

\subsection{Dynamics without pinning}
\label{sec.nopinning}

\begin{figure}[ht]
\begin{center}
\includegraphics[width=0.45\textwidth,clip]{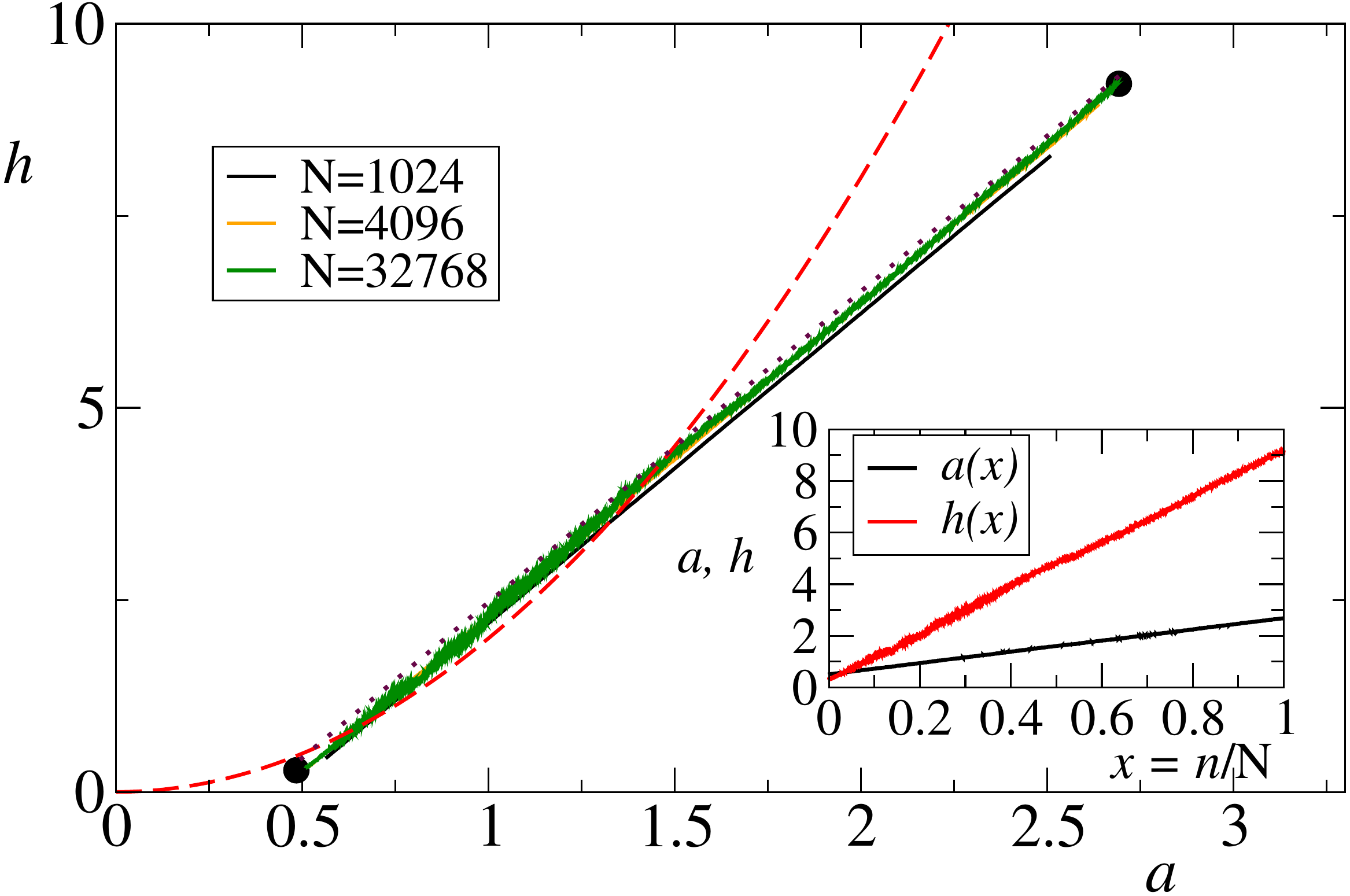}
\end{center}
\caption{
Unpinned dynamics.
 Parametric profiles of the unpinned C2C model for the same boundary parameters as in Fig~\ref{fig.c2c_pinned} and different system sizes.
The dotted purple line is a reference straight line. 
The red dashed line is the critical line, $h=2a^2$.
Inset: spatial mass  and energy  profiles for the largest size $N=32768$. 
For the largest size the stationary state was sampled for a time $\tau=2\times 10^8$ units.
}
\label{fig.c2c_unpinned_1}
\end{figure}

When pinning is removed, localized energy can diffuse much more efficiently in the system and
peaks can even attain the regions of the chain that are close to the
reservoirs, i.e. in the homogeneous phase.
This process has two important consequences.
Firstly, it introduces an effective dissipation mechanism for condensed energy, which prevents 
the unlimited growth of the condensate, thereby allowing to attain a
nonequilibrium stationary state at long times. This can be checked by comparing
\red{the incoming and outgoing  fluxes of mass and energy. For the two transported quantities we have separately 
verified that in the long time limit
$(|j^{(L)}| - |j^{(R)})|/(|j^{(L)}| + |j^{(R)}|) \rightarrow 0$, as expected
for stationary states.}
Secondly, peaks' diffusion smooths the density profiles and makes them linear, 
both in the parametric representation $h(a)$ (see Fig.~\ref{fig.c2c_unpinned_1}, main) and
when plotted versus space (lower inset).

\begin{figure}[ht]
\begin{center}
\includegraphics[width=0.49\textwidth,clip]{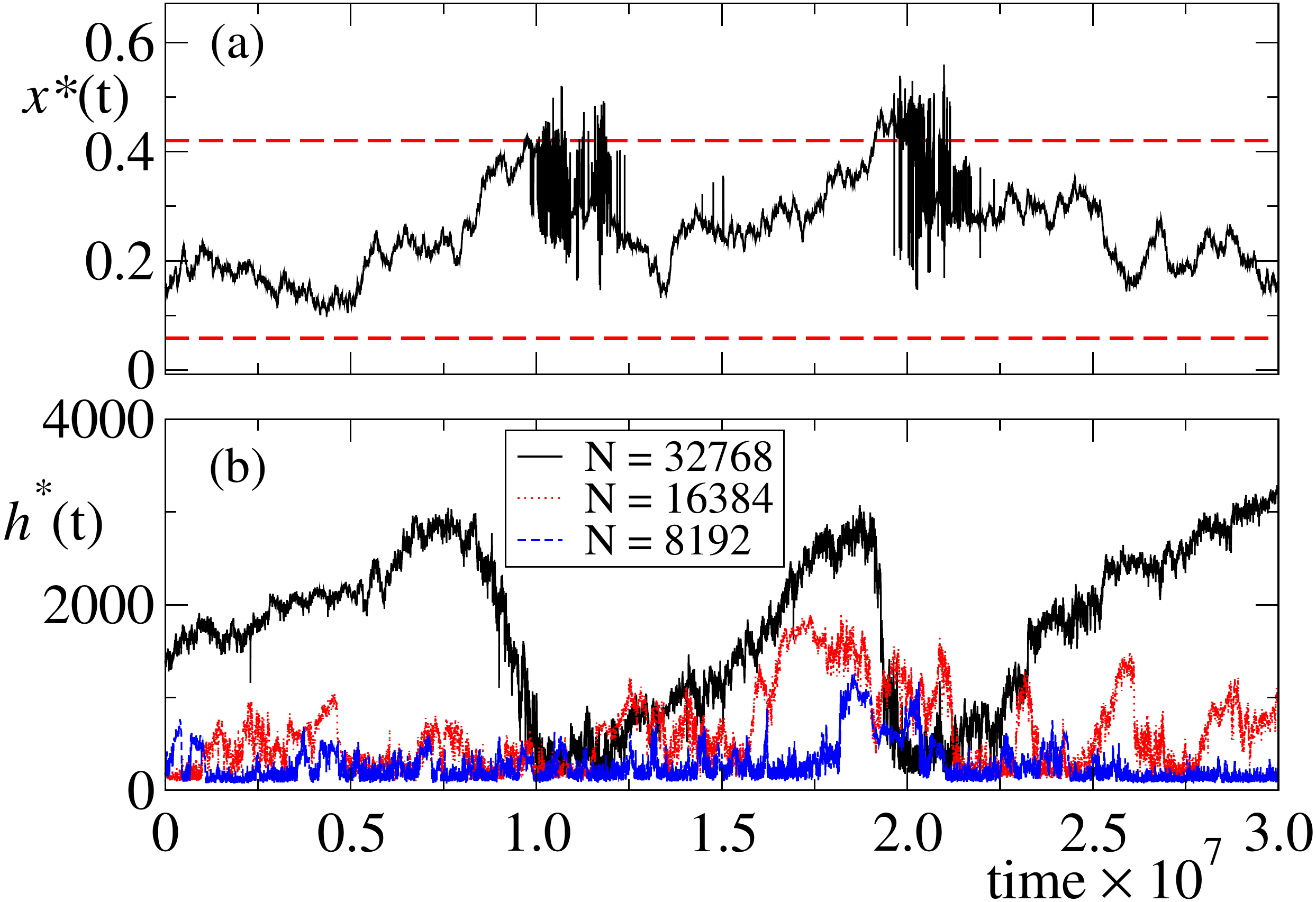}
\end{center}
\caption{
Unpinned dynamics.
Evolution  of the condensate position $x^*(t)\equiv n^*(t)/N$ for the largest size $N=32768$ and after the transient $\tau_0$.
The two  horizontal dashed lines identify the two positions where the stationary profile in (a) crosses the transition line $h=2a^2$. For each time, $n^*(t)$ is measured as the position of the site with largest energy along the chain. 
(c) Evolution of the condensate energy $h^*(t)=\e_{n^*(t)}$ on the same time scale of panel (b) and for different sizes.
}
\label{fig.c2c_unpinned_2}
\end{figure}

The nature of the dissipation mechanism and of the stationary state can be further clarified by monitoring the evolution of the instantaneous position
 and energy of the condensate, as shown in
Fig~\ref{fig.c2c_unpinned_2}.
Panel~(a) shows that the condensate position $n^*(t)$ (black line), identified here as the lattice site with the highest energy in the whole chain,
is approximately confined in the supercritical region $h>2a^2$, between the two red dashed lines. 
We interpret this confinement as the evidence that the condensate can not survive too close to the reservoirs.
Indeed the behavior of the condensate energy $h^*(t)$ (panel~(b)) confirms that when the condensate approaches the homogeneous
region (specifically the side of the hotter reservoir) it undergoes a decay process that eventually destroys it. Remarkably, after this event a new condensate is created, in agreement with the stationary nature of the process.
The sequence of growth and decay cycles is found to depend on the system size: larger $N$ produce slower cycles, with 
larger peaks (see different curves in panel~(b)).

\begin{figure}[ht]
\begin{center}
\includegraphics[width=0.45\textwidth,clip]{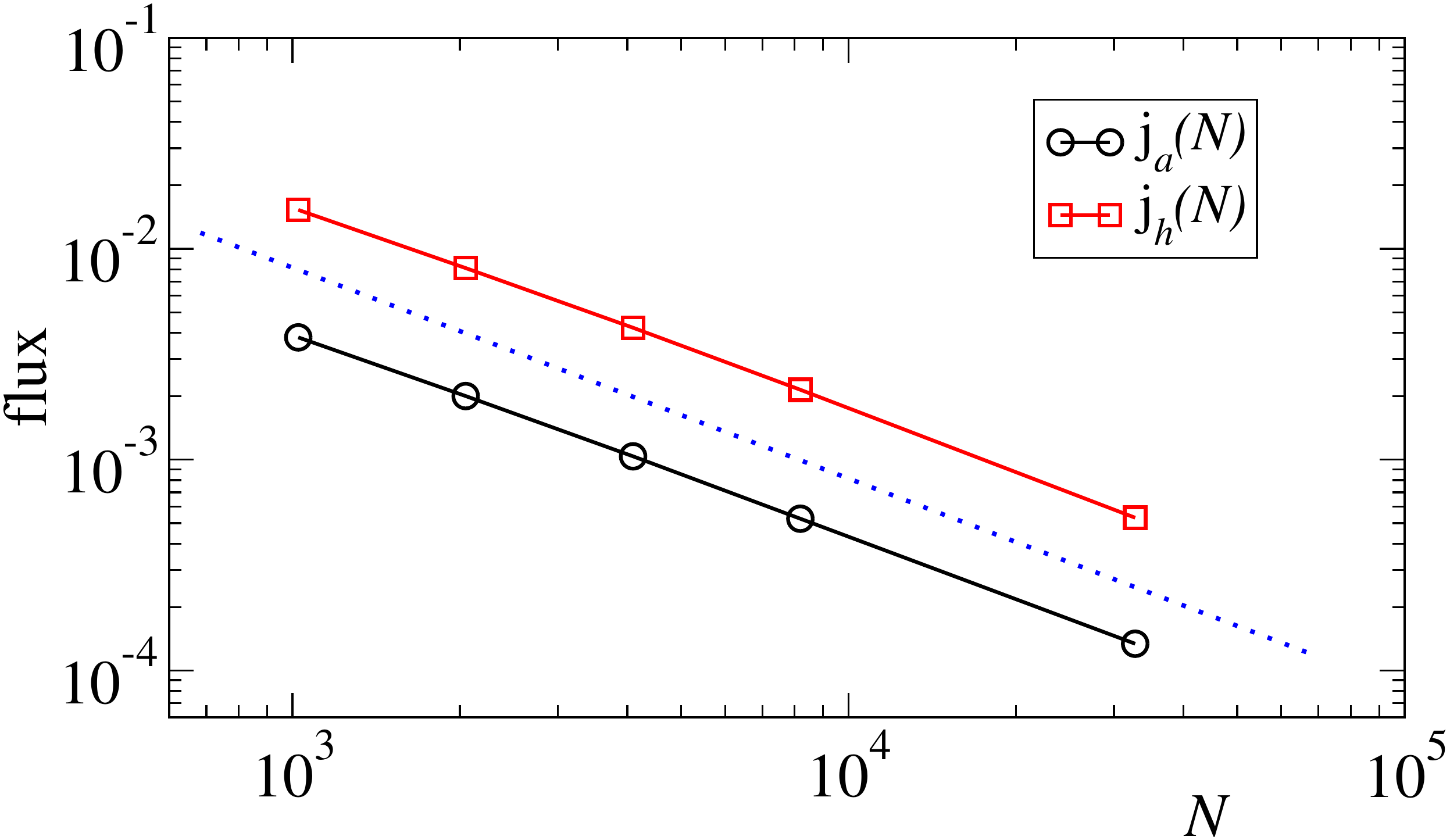}
\end{center}
\caption{
\red{Unpinned dynamics. Scaling of the stationary mass and energy fluxes {\it vs} $N$: the blue dotted line represents a diffusive scaling law $j\sim N^{-1}$.
}}
\label{fig.c2c_unpinned_3}
\end{figure}

\red{Finally, in Fig.~\ref{fig.c2c_unpinned_3} we report the behavior of the mass ($j_a$) and energy ($j_h$) currents 
as a function of the system size $N$ 
(we can now omit superscripts $^{(L)}$ and $^{(R)}$ because in a steady state left and right currents are equal).
We find the standard scaling $j_{a,h}\sim 1/N$,   
which  corresponds to normal, diffusive transport~\cite{Lepri_PhysRep}.
}

\section{Discussion}\label{sec:disc}
\red{The observation of condensation phenomena in equilibrium conditions  requires 
that the control parameter of the system exceeds a certain critical threshold. 
Such a condition has been realized in an isolated system initialized in
the condensed region of parameters or imposing an overcritical boundary condition. 
In this paper, we have given evidence that a new mechanism is able to produce condensation when {\it i)} the system is kept
out of equilibrium by boundary imbalances and {\it ii)} at least two independent quantities are transported (coupled transport).  
If taken individually, each boundary condition corresponds to a  thermal state at finite temperature and belongs to the
homogeneous, delocalized region. In fact, using 
standard heat baths like the Monte Carlo ones implemented here it would be impossible to
equilibrate the system in the localized region~\cite{Gradenigo2021_JSTAT}:
localization emerges here because it is a pure nonequilibrium effect!
The main question is therefore the role of the out-of-equilibrium setup in favoring
the condensation process.
}

To clarify this point we propose a simple argument based on the dynamical evolution of a triplet of sites
whose central site hosts a peak with large mass $b$. The two external sites of the triplet 
are supposed to be characterized by given mass distributions, \red{imposed by} the left and right reservoirs.
Given this simplified picture, \red{we analyze} if on average the dynamical move favors the
relaxation of the peak or on the contrary it may favor its growth.
Let us therefore consider the triplet of masses $(c_1,b,c_2)$ which evolves microcanonically towards
$(c_1 + x, b+ \delta b, c_2 +y)$ in such a way that mass and energy are conserved,
\bea
\label{eq.mc}
&& x+y+ \delta b = 0 \\
&& x^2 + y^2 + (\delta b)^2 + 2(c_1 x + c_2 y +b\delta b) = 0 .
\label{eq.ec}
\eea

In equilibrium conditions $c_1$ and $c_2$ are drawn from the same distribution because
boundary conditions are symmetric. In general it is convenient to write
$c_i = \bar c_i + \delta_i$, where $\bar c_i$ are the average values of the mass distributions
at the left ($i=1$) and at the right ($i=2$) of the breather.
From Eqs.~(\ref{eq.mc}-\ref{eq.ec}) simple algebra allows to find the exact relation
\be
\delta b (b-c) = -Q - (\delta_1 x + \delta_2 y) + \Delta(y-x),
\label{eq.triplet}
\ee
where $Q=\frac{1}{2}[x^2 + y^2 + (x+y)^2] \ge 0$ while $c$ and $\Delta$ are defined by the relations
$\bar c_1 = c + \Delta$ and  $\bar c_2 = c - \Delta$.
A value $\Delta \ne 0$ means that we are out-of-equilibrium.
Assuming $b > c$ because $b$ is a peak, a negative (positive) right-hand side of Eq.~(\ref{eq.triplet}) means 
that the condensate decreases (increases) its mass.
The first term on the right-hand side is negative-definite and it is at the origin of the condensate relaxation to equilibrium,
because the second term fluctuates around zero ($\bar\delta_i =0$ by definition) 
and the third term vanishes for symmetric boundary conditions.

In out-of-equilibrium conditions, $\Delta \ne 0$,
one finds that the transport of mass between the side with the largest $\bar c_i$ and the opposite side
determines that $(y-x)$ has on average the same sign as $\Delta$.
Therefore the out-of-equilibrium setup contributes to the variation of mass
of $b$ with a term which on average is positive.
This clarifies that boundary imbalances favors the development of
condensation.

We may now wonder about the relevance of having two conservation laws,
especially in light of the results of Refs.~\cite{Gradenigo2021_JSTAT,Gradenigo2021_EPJE}
where it is shown that the equilibrium properties of C2C model are equivalent to
a model where only one quantity is conserved, the energy $H=\sum_i \e_i$,
but where $\e_i \ge 0$ are independent variables following the distribution
$f_\lambda(\e) = (\lambda/2\sqrt{\e}) \exp(-\lambda \sqrt{\e})$.
For this model the critical value of the energy density $h=H/N$ above which condensation appears 
is $h_c = \langle\e\rangle =2/\lambda^2$~\cite{Godreche2001_EPJB,Evans2006_JSP}.
If we attach such a lattice model to reservoirs at different temperatures, only a single current
(the heat flux) can flow,
and the resulting temperature profile must be monotonic~\cite{lepri2016}. Therefore, if
boundary conditions are subcritical, i.e. they do not favor localization,
the same must be true throughout the system.
The above scenario is further confirmed by the study of the Zero-Range Process with open, asymmetric boundary conditions~\cite{Levine2005}.
This is a model with only one conservation law and authors find localization only if boundary conditions
allow it at equilibrium. It is also worth mentioning that in such case the condensate appears
in proximity of the overcritical boundary.

\section{Perspectives}
\label{sec.perspectives}
As a future perspective of this work, coupled transport phenomenology of the C2C
model will be worthy of complementary analysis from the point of view of linear response theory~\cite{livi17}.
The main open questions include how the Onsager transport coefficients 
vary in the parameter space $(a,h)$ and, in particular, whether they display any peculiar
behavior when  parametric curves cross the critical line $h=2a^2$.
It will also be interesting to understand in more detail how energy peaks are born and die
in the unpinned case. In fact, these processes appear to be relevant both for the appearance of 
a nonequilibrium stationary state and of the linearity of mass and energy profiles, 
 see the inset in Fig.~\ref{fig.c2c_unpinned_1}.

In light of the importance of conservation laws it will be useful to study the observed condensation process in other models
with conserved quantities. Among them, the DNLS equation is undoubtedly a relevant model to investigate.
In fact, a closely related phenomenology  was found in ~\cite{Iubini2017_Entropy}  when a DNLS chain
was attached to a mass dissipator to one boundary and to a reservoir at the opposite boundary. 
A mass dissipator is not a proper reservoir, as it drives
the system towards the point $(a,h)=(0,0)$, which is singular thermodynamically, because
any isothermal curve, including $T=0$ and $T=\infty$, terminates there, see Fig.~\ref{fig.ahTmu}(a).
We also remark that despite the DNLS and C2C models display very similar equilibrium, critical properties, they have extremely
different relaxation dynamics~\cite{PRL_DNLS,PPI2022_JSTAT}: 
the frozen dynamics of localized states is a hallmark of the Hamiltonian dynamics of the DNLS equation
 and makes practically impossible
to attain equilibrium in the condensed region.
Therefore, the behavior of the DNLS model in the out-of-equilibrium setup studied here is an open question.

It would also be interesting to study other stochastic models where the local energy is not quadratic
with the mass. In practice they are defined as a generalization of the C2C model where $A=\sum_i c_i$ is unchanged and $H =\sum_i c_i^\alpha$. 
These models have been proposed in Ref.~\cite{Szavits2014_PRL} and they are known to display condensation
at equilibrium~\cite{Mori2021}. Their relaxation dynamics and nonequilibrium properties are not known.

\red{More generally it would be useful to find suitable protocols for driving a system from the homogeneous
to the localized phase. This would be relevant in some experimental contexts where
the preparation of a system in the condensed region might be difficult  because of dynamic or thermodynamic reasons~\cite{Baldovin_2021_review}.	
}

\begin{acknowledgments}
Authors thank Antonio Politi for discussions and Federico Corberi for a critical reading of
a first draft of the manuscript. 
PP acknowledges support from the MIUR PRIN 2017 project 201798CZLJ.
\end{acknowledgments}

\appendix

\section{The C2C model in the grand canonical ensemble}
\label{app:gcan}
In this Appendix we derive explicitly the equilibrium properties of the C2C model within the grand canonical 
ensemble.

Given the two conserved quantities $A$ and $H$ in Eq.~(\ref{eq:AH}),
the grand canonical partition function is 
\bea
Z(\beta,\mu) &=& \int_0^\infty dc_1\dots dc_N\;  e^{-\beta (H-\mu A)} \nonumber\\
&=&
\int_0^\infty dc_1\dots dc_N\;  e^{-\beta (\sum_n c_n^2-\mu \sum_n c_n)} \nonumber\\
&=&
\prod_{n=1}^N \int_0^\infty  dc_n \, e^{-\beta (c_n^2-\mu c_n)}\nonumber \\
&\equiv & Z_0^N (\beta,\mu)
\eea
where
\be
Z_0(\beta,\mu) = 
\frac{1}{2} \sqrt{\frac{\pi}{\beta}} e^{\beta\mu^2/4}
\left[ 1 + \erf\left(\frac{\sqrt{\beta}\mu}{2}\right)\right] .
\ee

The mass density $a$ and the energy density $h$ can be found from the equations
\bea
a &=& \frac{\langle A\rangle}{N}  = \frac{1}{\beta} \frac{\partial}{\partial\mu} \ln Z_0 \\
h &=& \frac{\langle H\rangle}{N} = - \frac{\partial}{\partial\beta} \ln Z_0 + \mu a
\eea
which give Eqs.~(\ref{eq.a_gc},\ref{eq.h_gc}).

The curve corresponding to a vanishing temperature is easily found. In fact, for $\beta\to +\infty$
Eq.~(\ref{eq.a_gc}) gives $a=\mu/2$ and Eq.~(\ref{eq.h_gc}) gives $h=a\mu/2 = a^2$.
The region $h<a^2$ is not allowed as clearly shown by the evaluation of the variance of the mass
distribution,
\be
\sigma^2 \equiv \bar{c^2} - (\bar c)^2 = h -a^2 \ge 0 .
\ee 

The curve corresponding to a diverging temperature $(\beta\to 0)$ is more subtle because
the first term on the right-hand side of Eq.~(\ref{eq.h_gc}) diverges.
In the same limit Eq.~(\ref{eq.a_gc}) would give an unphysical diverging mass if $\mu$
keeps finite. It is therefore necessary that $\mu$ diverges as well and the correct
divergence is such that $\beta\mu = -\gamma$, where $\gamma$ is a finite, positive constant.
Expanding the $\erf$ function in Eq.~(\ref{eq.a_gc}) we find
\be
a \simeq -\frac{1}{\beta\mu} + \frac{4}{\beta^2\mu^3} 
\ee
and Eq.~(\ref{eq.h_gc}) finally gives
\be
h = \frac{2}{\beta^2\mu^2} = 2a^2 .
\ee

\section{Dynamics of the C2C model}
\label{app:mmc}

A Monte Carlo time step is composed by a sequence of $N$ attempted moves,
where $N$ is the number of sites. 
An attempted move starts choosing a random site $k=1,\dots,N$. If $k=1$ or $k=N$
we implement the heat baths as explained here below.
If $k\ne 1,N$ we consider the triplet of consecutive sites $(k-1,k,k+1)$ and
we perform a microcanonical move of the masses $(c_{k-1},c_k,c_{k+1})$,
i.e. a transition towards a new triplet $(c'_{k-1},c'_k,c'_{k+1})$
having the same mass and energy,
\bea
\label{eq.c2cM}
c_{k-1} + c_k + c_{k+1} = &M& = c'_{k-1} + c'_k + c'_{k+1} \\
c^2_{k-1} + c^2_k + c^2_{k+1} = &E& = (c'_{k-1})^2 + (c'_k)^2 + (c'_{k+1})^2
\label{eq.c2cE}
\eea 
and chosen uniformly in the solution space of previous equations.

In practice, Eqs.~(\ref{eq.c2cM}-\ref{eq.c2cE}) define the intersection between
a plane and a sphere, with the constraint that all new masses are positive (see Fig.~\ref{fig.c2c}).
This intersection (which is non-zero because there is the solution $c'_i = c_i$,
$i=k-1,k,k+1$) is either a circle (if all $c_i$ are of the same order) 
or the sum of three disjoint arcs of a circle (if one $c_i$ is much larger
than the other two).
Solutions can be parameterized by an angle whose
random choice ensures that detailed balance is locally satisfied.

We have implemented Monte Carlo heat baths attached to the lattice sites $k=1$ and $k=N$ as follows.
Once that one of the two external sites is randomly chosen we perform an attempt of
variation of its mass, 
\be
c_k \to c'_k = c_k + \psi,
\ee
where $\psi$ is a random variable  uniformly distributed in the interval $(-\psi_k,\psi_k)$.
If $c'_k \ge 0$ we accept the move according to the Metropolis rule with the cost function
\be
W=e^{-(\Delta H -\mu_k\Delta A)/T_k},
\ee 
where $(T_k,\mu_k)$ are the temperatures and the 
chemical potentials of the heat baths, $\Delta H = (c_k + \psi)^2 - c_k^2$,
and $\Delta A = \psi$.
\red{We have tested several values of $\psi_1$ and $\psi_N$ and we generally used
$\psi_1=\psi_N=0.5$.}

\end{document}